\documentclass[camera]{jpaper}

\usepackage[nocompress]{cite}
\usepackage{algorithmic}
\usepackage{array}

\makeatletter
\let\MYcaption\@makecaption
\makeatother

\usepackage[font=footnotesize]{subcaption}

\makeatletter
\let\@makecaption\MYcaption
\makeatother

\usepackage{fixltx2e}
\usepackage{dblfloatfix}
\usepackage[nolessnomore, italic]{mathastext}
\usepackage[T1]{fontenc}
\usepackage[usenames,dvipsnames,svgnames,table]{xcolor}
\usepackage[normalem]{ulem}
\usepackage{enumitem}
\usepackage{setspace}
\usepackage{indentfirst}
\usepackage{footmisc}
\usepackage{fancyhdr}
\usepackage{authblk}
\usepackage[us,12hr]{datetime}
\usepackage[keeplastbox]{flushend}

\usepackage{wrapfig}    
\usepackage{rotating} 

\usepackage{amssymb}    

\usepackage{stmaryrd}

\usepackage{bbding}

\usepackage{comment}
\usepackage{hhline}
\usepackage{array}

\usepackage{xspace}
\usepackage{soul}
\usepackage{graphicx}
\usepackage{url}
\usepackage{caption}
\usepackage{fancyvrb}
\usepackage{multirow}

\usepackage{algorithm}
\usepackage{algorithmic}
\usepackage{wrapfig}
\usepackage{tikz}
\usepackage{amsmath}
\usepackage{booktabs}

\newif\ifcameraready
\camerareadytrue

\newcommand{\versionnum}[0]{5.1}

\ifcameraready
\else
\fi

\ifcameraready
  \newcommand{\todo}[1][]{}
\else
  \newcommand{\todo}[1][]{\textbf{\fcolorbox{black}{red}{\color{white}{TODO}}} \underline{$\overline{\hbox{\emph{#1}}}$}}
\fi

\definecolor{lightgreen}{RGB}{195, 233, 211}
\newcommand\yellow{\bgroup\markoverwith
  {\textcolor{lightgreen}{\rule[-.5ex]{.1pt}{2.5ex}}}\ULon}
\sethlcolor{lightgreen}

\definecolor{darkgreen}{RGB}{70,168,70}

\newboolean{publicversion}
\setboolean{publicversion}{true}

\ifthenelse{\boolean{publicversion}}{
  \pagenumbering{arabic}
  \newcommand{\grumbler}[2]{}
  \newcommand{\assign}[1]{}
  \newcommand{\respond}[3]{}
  \newcommand{\changesI}[0]{}

  \newcommand{\sg}[0]{}

}{
  \pagenumbering{arabic}
  \newcommand{\grumbler}[2]{\textcolor{blue}{\bf #1: #2}}
  \newcommand{\assign}[1]{\textcolor{purple}{\bf RESPONSIBLE: #1}}
  \newcommand{\respond}[3]{\textcolor{#1}{\bf #2-response: #3}}
  \newcommand{\changesI}[1]{\textcolor{MidnightBlue}{#1}}

  \newcommand{\sg}[1]{\textcolor{BrickRed}{#1}}
}

\newcommand{\cmdact}{\texttt{ACTIVATE}\xspace}
\newcommand{\cmdacts}{\texttt{ACTIVATEs}\xspace}
\newcommand{\cmdrd}{\texttt{READ}\xspace}

\newcommand{\cmdpre}{\texttt{PRECHARGE}\xspace}

\newcommand{\cmdsel}{\texttt{SA\_SEL}\xspace}

\newcommand{\authspace}[0]{\qquad}
\newcommand{\affilspace}[0]{\qquad}

\begin{document}
%
\title{Exploiting the DRAM Microarchitecture\\to Increase Memory-Level Parallelism}


\author{
    Yoongu Kim$^{1}$\authspace%
    Vivek Seshadri$^{2,1}$\authspace%
    Donghyuk Lee$^{3,1}$\authspace%
    Jamie Liu$^{4,1}$\authspace%
    Onur Mutlu$^{5,1}$}%
\affil{{\it%
	$^1$Carnegie Mellon University\affilspace%
	$^2$Microsoft Research India}%
\vspace{2pt}\\{\it%
	$^3$NVIDIA Research\affilspace%
	$^4$Google\affilspace%
    $^5$ETH Z{\"u}rich}}%

\maketitle

\begin{abstract}

This paper summarizes the idea of 
Subarray-Level Parallelism (SALP) \changesI{in DRAM}, which was published in ISCA 2012~\cite{salp}, \changesI{
and examines the work's significance and future potential}.  Modern DRAMs have
multiple banks to serve multiple memory requests in parallel.  However, when
two requests go to the same bank, they have to be served serially, exacerbating
the high latency of off-chip memory. Adding more banks to the system to
mitigate this problem incurs high system cost. Our goal in this work is to
achieve the benefits of increasing the number of banks with a low-cost
approach. To this end, we propose three new mechanisms,
SALP-1, SALP-2, and MASA (Multitude of Activated Subarrays), to reduce the
serialization of different requests that go to the same bank. The key observation
exploited by our mechanisms is that a modern DRAM bank is implemented as a
collection of \emph{subarrays} that operate largely independently while sharing few
global peripheral structures.

Our three proposed mechanisms mitigate the negative impact
of bank serialization by overlapping different components of the bank access
latencies of multiple requests that go to different subarrays within the same
bank. SALP-1 requires no changes to the existing DRAM structure, and needs to only
reinterpret some of the existing DRAM timing parameters. SALP-2 and MASA require only
modest changes ($<$ 0.15\% area overhead) to the DRAM peripheral structures,
which are much less design constrained than the DRAM core.  
Our evaluations show that SALP-1, SALP-2 and MASA significantly improve
performance for both single-core systems (7\%/13\%/17\%) and multi-core systems
(15\%/16\%/20\%), averaged across a wide range of workloads.  We also
demonstrate that our mechanisms can be combined with application-aware memory
request scheduling in multi-core systems to further improve performance and
fairness. 

\changesI{Our proposed technique has enabled significant research in
the use of subarrays for various purposes \sg{(e.g., \cite{seshadri2013rowclone,tl-dram,
dsarp,ava-dram,lisa,chargecache,choi-isca2015,cream,lu-micro2015,yue-date13,
son-isca2013, ambit, seshadri-cal2015, lym.hpca18})}.
SALP has also been described and
evaluated by a recent work by Samsung and Intel~\cite{kang14} as a promising
mechanism to tolerate long write latencies that are a result of aggressive DRAM
technology scaling.}
\end{abstract}


\section{Introduction}

\label{sec:problem}

To be able to serve multiple memory requests in parallel, modern DRAM chips
employ multiple {\em banks} that can be accessed independently, providing \emph{bank level
parallelism}. Unfortunately,
if two memory requests go to the same bank, they have to be served one after
another. This is \changesI{called} a {\em bank conflict}. In the worst
case, bank conflicts may delay a memory request by hundreds or even thousands of
nanoseconds~\cite{salp,seshadri2013rowclone,chargecache,lisa}. In particular, bank conflicts cause three specific
problems that degrade the access latency, bandwidth utilization, and
energy efficiency of the main memory subsystem:

\begin{enumerate}
  \item \textbf{Serialization.} Bank conflicts serialize requests that could
    potentially have been served in parallel. Such serialization exacerbates the
    already large latency of a memory access, and may cause processor cores to stall
    for much longer.
  \item \textbf{Write Recovery.} A request scheduled after a write request to the
    same bank experiences an extra delay called the \emph{write recovery penalty},
    which is an additional time required to safely store new data in the cells.
    This write recovery latency further aggravates the impact of serialization.
  \item \textbf{Row Buffer Thrashing.} Each bank has a \emph{row buffer} that caches
    the last accessed row. A request that hits in the row buffer is much cheaper
    in terms of both latency and energy than a request that misses in the row buffer. 
    However, bank conflicts between requests that access different rows lead to 
    costly row buffer misses.
\end{enumerate}

A naive solution to bank conflicts is to increase the
number of banks. Unfortunately, as we discuss in Section~1 of our ISCA 2012 paper~\cite{salp},
simply adding more banks to the memory subsystem
comes at significantly high costs or reduced performance 
regardless of the way it is done: more
banks per chip, more ranks per channel, or more channels.\footnote{We refer the reader to our prior 
works\changesI{~\cite{atlas,tcm,salp,tl-dram,al-dram,chargecache,raidr,dsarp,lisa,seshadri2013rowclone,chang-sigmetrics17,
ava-dram,ambit,liu-isca2013, chang-sigmetric16, softmc, kim-cal2015, donghyuk-ddma, 
donghyuk-stack, kim-isca2014, patel.isca17, kim.hpca18}} for a detailed background on DRAM.}




The goal in our ISCA 2012 paper~\cite{salp} is to mitigate such detrimental
effects of bank conflicts in a cost-effective manner. \changesI{Toward} that end, we make
two key observations that lead to our proposed solutions.

\begin{figure}[h!]
\centering
\begin{subfigure}{0.45\linewidth}
\centering
\includegraphics[height=1.1in]{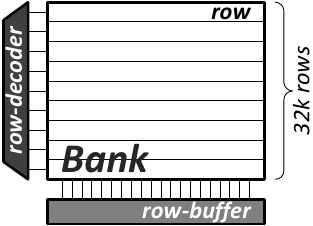}
\caption{Logical abstraction}\label{fig:bank-abstraction}
\end{subfigure}
\begin{subfigure}{0.53\linewidth}
\centering
\includegraphics[height=1.1in]{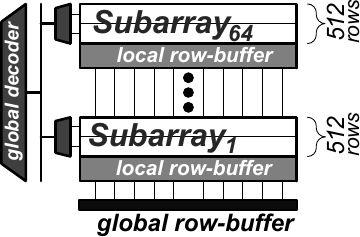}
\caption{Physical implementation}\label{fig:bank-implementation}
\end{subfigure}

\caption{DRAM bank organization. \changesI{Adapted} from~\cite{salp}.}
\label{fig:bank-abstraction-implementation}
\end{figure}

{\bf Observation 1.} A modern DRAM bank is \changesI{\emph{not}} implemented as a monolithic
component equipped with only a single row buffer. Implementing a DRAM bank in
such a way requires very long internal wires (called \emph{bitlines}) to connect the
row buffer to all the rows in the bank, which can significantly increase the
access latency. Instead, as Figure~\ref{fig:bank-abstraction-implementation}b
shows, a bank consists of multiple {\em subarrays}, each with its own {\em local
row buffer}.  Subarrays within a bank share two important global structures:
{\em i)} a {\em global row address decoder,} and {\em ii)} a {\em global
row buffer}.

\begin{figure*}[ht!!!]
\centering
\includegraphics[width=0.95\linewidth]{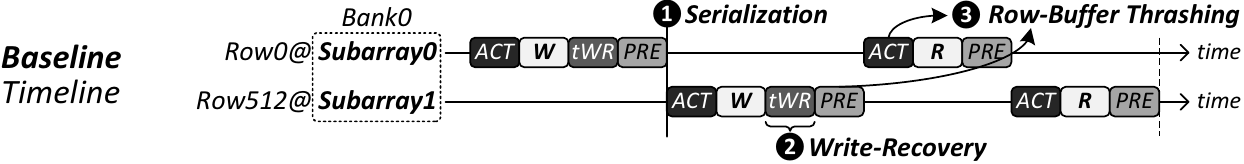}
\caption{Timeline of four requests to two different rows in the same bank. Adapted from~\cite{salp}.}
\label{fig:timeline-baseline}
\end{figure*}

\begin{figure*}[ht!!!]
\centering
\includegraphics[width=0.95\linewidth]{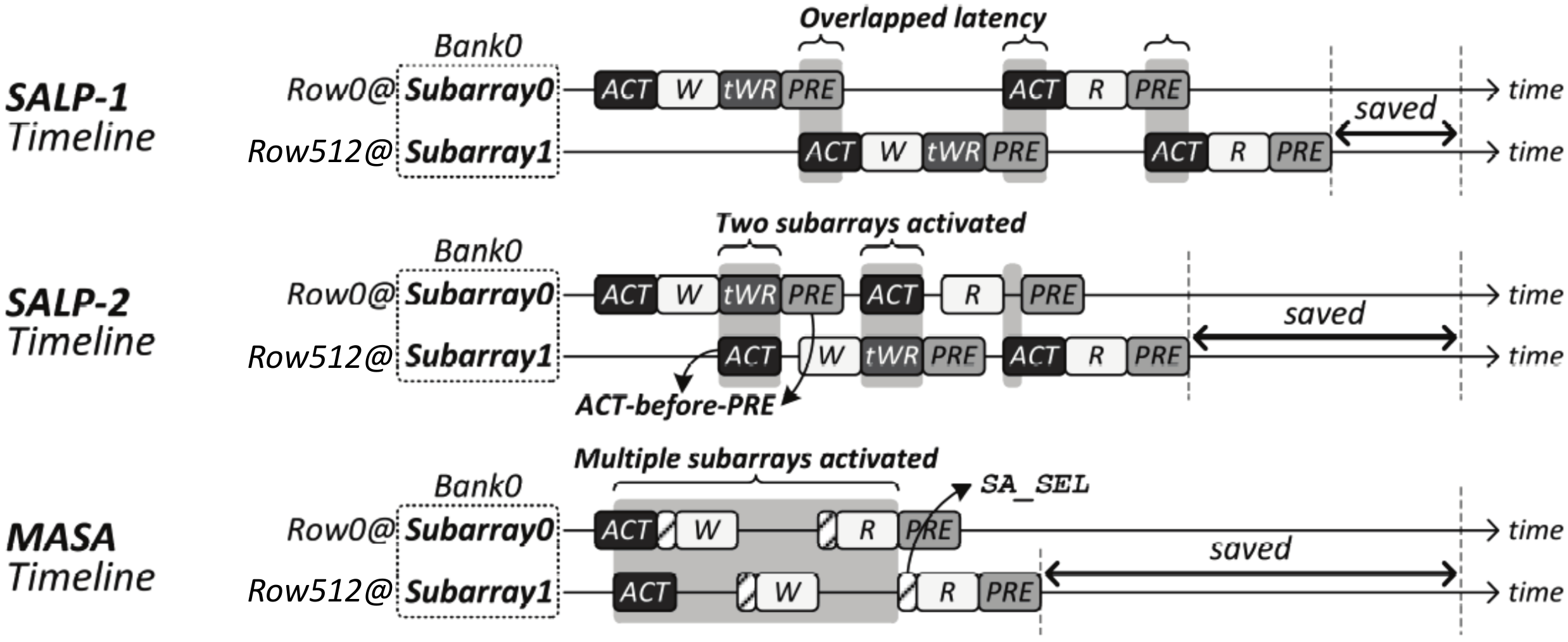}
\caption{Timeline of four requests to two different rows in the same bank but
  different subarrays, using our mechanisms to exploit subarray-level parallelism. Adapted from~\cite{salp}.}
\label{fig:timeline-salp}
\end{figure*}

{\bf Observation 2.} The latency of a bank access predominantly consists of three major
components: {\em i)} loading a row into the local row buffer ({\em activation}),
{\em ii)} accessing the data from the local row buffer ({\em read} or {\em
write}), and {\em iii)} clearing the local row buffer ({\em precharging})~\cite{ava-dram,softmc,chargecache,chang-sigmetric16,salp,al-dram,tl-dram}. 
In existing DRAM banks, all three operations must be completed for one request
before serving another request to a different row, even if the two rows reside
in \changesI{\emph{different}} subarrays. However, this does not need to be the case for two reasons. First,
activation and precharging are mostly local to each subarray, which enables the
opportunity to overlap these operations when they are to different subarrays.
Second, if we reduce the sharing of the global structures among subarrays, we
can \emph{parallelize} the concurrent activation of different subarrays. Doing so would
allow us to exploit the existence of \changesI{\emph{multiple}} local row buffers across the
subarrays, enabling more than just a single row to be cached for each bank and
thereby increasing the row buffer hit rate.

\section{Subarray-Level Parallelism}

{\bf Subarray-Oblivious Baseline.} Let us consider the baseline example shown
in Figure~\ref{fig:timeline-baseline}, which presents a timeline of four
memory requests being served at the same bank in a subarray-oblivious manner.
This example highlights the three key problems that we discussed in
Section~\ref{sec:problem}. First, requests are completely serialized, even
though they are to different subarrays. Second, although the write-recovery
penalty is local to a subarray, it delays a subsequent request to a different
subarray. Third, a request to one subarray unnecessarily evicts (i.e.,
\emph{precharges}) the other subarray's local row buffer, which must be reloaded
(i.e., activated) when a future request accesses the evicted row.  In this
section, we describe how SALP-1, SALP-2 and MASA can take an advantage of the
DRAM bank organization to enable parallel DRAM operations in a cost-effective
manner.

\subsection{SALP-1: Subarray-Level-Parallelism-1}

We observe that precharging and activation are mostly
local to a subarray. 
Based on this observation, we propose SALP-1, which overlaps the precharging
of one subarray with the activation of another subarray. In contrast, existing
systems always serialize precharging and activation to the same bank,
conservatively provisioning for when they are to the same subarray.  SALP-1
requires \emph{no modifications} to existing DRAM structure. It only requires
reinterpretation of an existing timing constraint (tRP) and, potentially, the
addition of a new timing constraint (which we describe in
Section~5.1 of our ISCA 2012 paper~\cite{salp}). Figure~\ref{fig:timeline-salp} (top)
shows the timeline of the same four requests from Figure~\ref{fig:timeline-baseline}
when we use SALP-1 instead of our Baseline.
As the timeline shows, overlapping the precharge operation
reduces the overall time needed to complete the four requests.

\subsection{SALP-2: Subarray-Level-Parallelism-2}

While SALP-1 pipelines the precharging and activation of different subarrays,
the relative ordering between the two commands is still preserved. This is
because existing DRAM banks do \changesI{\emph{not}} allow two subarrays to be activated \emph{at the
same time}. As a result, the write-recovery latency
of an activated subarray \changesI{delays not only
a \cmdpre to itself, but also a} subsequent \cmdact to another subarray.
Based on the observation that the \changesI{write-recovery} latency is also local to a
subarray, we propose SALP-2.  SALP-2 issues the \cmdact to another subarray
{\em before} the \cmdpre to the currently-activated subarray. As a result,
SALP-2 can overlap the write recovery of the currently-activated subarray with
the activation of another subarray, further reducing the service time compared
to SALP-1 (as shown in the middle timeline of Figure~\ref{fig:timeline-salp}).

However, as highlighted in the figure, SALP-2 requires two subarrays
to remain activated at the same time. This is not possible in existing
DRAM banks as the global row-address latch, which determines the
wordline in the bank that is raised, is shared by all of the
subarrays. Section~5.2 of our ISCA 2012 paper~\cite{salp} discusses how
to enable SALP-2 by eliminating this sharing. \changesI{The key idea is to
push the global address latch to each subarray, thereby creating local address
latches, one per subarray.}

\subsection{MASA: Multitude of Activated Subarrays}


Although SALP-2 allows two subarrays within a bank to be activated, it requires
the controller to precharge one of them before issuing a column command (e.g.,
\cmdrd) to the bank. This is because when a bank receives a column command, all
activated subarrays in the bank will connect their local row buffers to the
global bitlines. If more than one subarray is activated, this will result in a
short circuit. As a result, SALP-2 \changesI{cannot} allow multiple subarrays to
concurrently remain activated and serve column commands.

To \changesI{solve} this, we propose MASA, whose key idea is to allow {\em multiple} subarrays
to be activated at the same time, while allowing the memory controller to
\emph{designate} exactly one of the activated subarrays to drive the global
bitlines during the next column command. MASA has two advantages over
SALP-2. First, MASA overlaps the activation of different subarrays within a
bank. Just before issuing a column command to any of the activated subarrays,
the memory controller \emph{designates} one particular subarray whose row buffer
should serve the column command. Second, MASA eliminates extra \cmdacts to the
same row, thereby mitigating row buffer thrashing. This is because the local
row buffers of multiple subarrays can remain activated at the same time without
experiencing collisions on the global bitlines. As a result, MASA further
improves performance compared to SALP-2, as shown in the
bottom timeline of Figure~\ref{fig:timeline-salp}.

{\sloppypar{\bf MASA: Overhead.} 
To designate one of the multiple activated
subarrays, the controller needs a new command, \cmdsel ({\em subarray-select}).
In addition to the changes required by SALP-2, MASA requires a single-bit latch
per subarray to denote whether a subarray is \emph{designated} or not. 
According to our detailed circuit-level analysis, MASA
increases the DRAM die-size by only 0.15\% (due to extra latches) and the static
power consumption by only $\sim$1\% (each additional activated subarray consumes
0.56{\it mW}). Also, the memory controller needs less than 256 bytes to track
the status of subarrays across all DRAM banks. We discuss a detailed implementation
of MASA, along with its overhead, in Section~5.3 of our ISCA 2012 paper~\cite{salp}.}

\section{\sg{Experimental} Methodology}

We evaluate our three mechanisms for subarray-level parallelism using Ramulator~\cite{kim-cal2015, ramulator-src}, an open-source 
cycle-accurate DRAM simulator that we developed which accurately models DRAM subarrays.
We use Ramulator as part of a cycle-level
in-house x86 multi-core simulator, whose front-end is based on Pin~\cite{pin}.
We calculate DRAM dynamic energy consumption by associating an energy cost with
each DRAM command, derived using Micron's DDR3 DRAM tool~\cite{micron-power}, Rambus'
DRAM power model~\cite{rambus-power}, and previously published data~\cite{dram-energy}. 

We evaluate SALP-1, SALP-2, and MASA on a wide variety of
workloads~\cite{tpc,spec2006,stream,gups} and system
configurations~\cite{intel-nehalem,intel-sandybridge,opensparc-t1,ibm-power7}.
The results shown in Section~\ref{sec:eval} are based on the conservative
assumption that a DRAM bank exposes only 8~subarrays to be exploited by our subarray-level
parallelism mechanisms,
whereas in practice the number of subarrays in current DRAM banks is typically
much higher ($\sim$64).  Section~9.2 of our ISCA 2012 paper~\cite{salp} shows
that the performance improvement of our three mechanisms over a subarray-oblivious
baseline increases with a greater number of subarrays.

For our full methodology, we refer the reader to Section~8 of our ISCA 2012 paper~\cite{salp}.


\section{\sg{Evaluation}}
\label{sec:eval}

\begin{figure*}[t]
\centering
\includegraphics[width=0.95\linewidth]{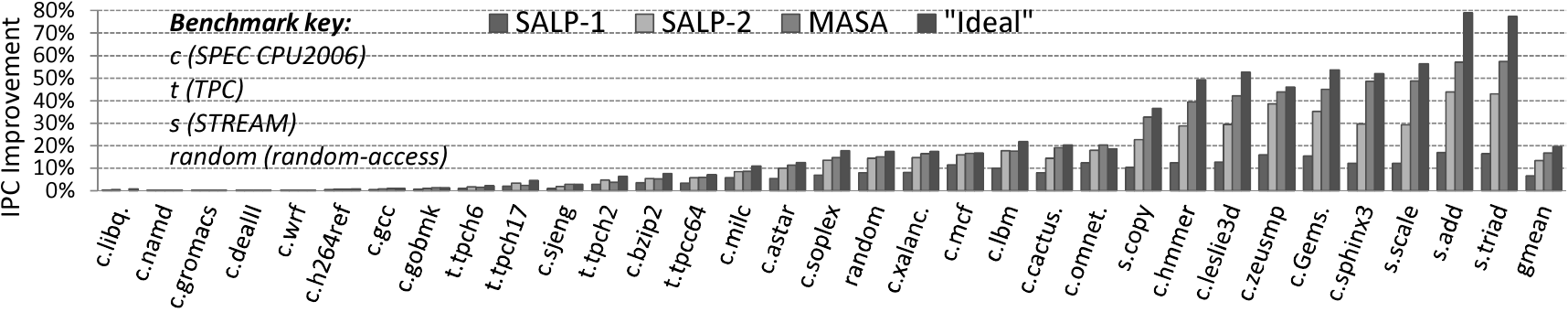}

\caption{IPC improvement of SALP-1, SALP-2, MASA, and an ideal mechanism over the subarray-oblivious baseline. Reproduced from~\cite{salp}.}

\label{bar:perwkld_ipc}
\end{figure*}

Figure~\ref{bar:perwkld_ipc} shows the performance improvement of SALP-1,
SALP-2, and MASA on a system with 8 subarrays-per-bank over a subarray-oblivious
baseline. The figure also shows the performance improvement of an ``Ideal''
scheme which is the subarray-oblivious baseline with 8 times as many banks (this
represents a system where all subarrays are fully independent). 
The benchmarks are sorted along the x-axis by increasing memory intensity.
We make two
observations from the figure. First, SALP-1, SALP-2, and MASA consistently perform better than the
baseline for all benchmarks. On average, they improve the average performance by 6.6\%,
13.4\%, and 16.7\%, respectively. Second, MASA captures most of the benefits of
``Ideal,'' which improves performance by 19.6\% compared to baseline.

The difference in performance improvement across benchmarks can be explained by
a combination of three factors related to \changesI{the benchmarks'} individual memory access behavior.
First, subarray-level parallelism in general is most beneficial for
memory-intensive benchmarks that frequently access memory (e.g., the benchmarks located
towards the right of Figure~\ref{bar:perwkld_ipc}). By increasing the memory
throughput for such applications, subarray-level parallelism significantly
alleviates their memory bottleneck. The average memory intensity of the
applications that gain $>$5\% performance with SALP-1 is
18.4 MPKI (last-level cache misses per kilo-instruction), compared to 1.14 MPKI
for the other applications.

Second, the advantage of SALP-2 is large for applications that are
write-intensive (i.e., those with the most write misses per kilo-instruction,
or WMPKI). For such applications, SALP-2 can overlap the long
write-recovery latency with the activation of a subsequent access. In
Figure~\ref{bar:perwkld_ipc}, the three applications that improve more than
38\% with SALP-2 are among both the most memory-intensive ($>$25 MPKI) and the
most write-intensive ($>$15 WMPKI).

Third, MASA is beneficial for applications that experience frequent bank
conflicts. For such applications, MASA parallelizes accesses to different
subarrays by concurrently activating multiple subarrays (\cmdact) and allowing
the application to switch between the activated subarrays at low cost (\cmdsel).
Therefore, the subarray-level parallelism offered by MASA can be gauged by the
\cmdsel-to-\cmdact ratio. For the nine applications that benefit more than 30\%
from MASA, on average, one \cmdsel was issued for every two \cmdacts, compared
to one-in-seventeen for all other applications. For a few benchmarks, MASA
performs slightly worse than SALP-2. This is because the baseline scheduling algorithm used with
MASA tries to overlap as many \cmdacts as possible, and in the process
inadvertently delays the column command of the most critical request.  This delay
to the most critical request
slightly degrades performance for these benchmarks.\footnote{For one benchmark,
MASA performs slightly better than ``Ideal'' due to interactions with the
scheduler.}

\begin{figure}[h]
\centering
    \includegraphics[height=0.9in]{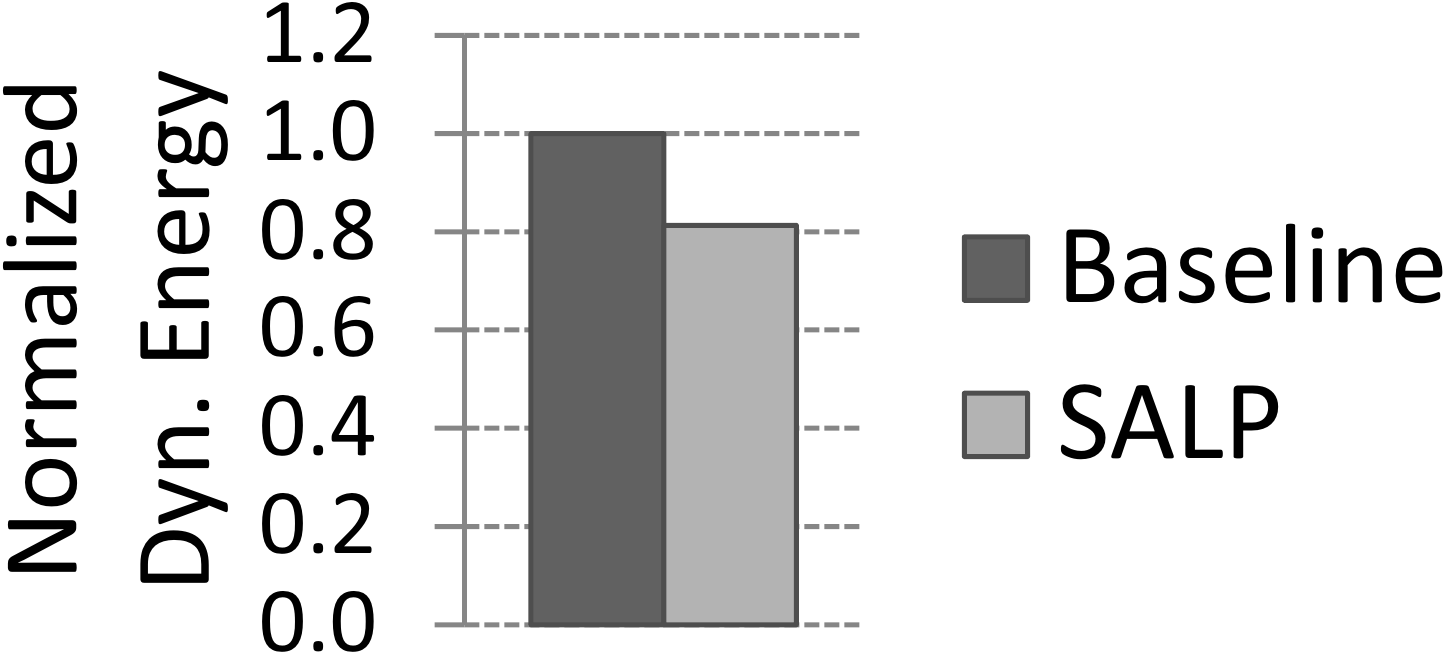}
\caption{Dynamic DRAM energy consumption for MASA.}
\label{bar:row}
\end{figure}

\sloppypar{{\bf Energy Efficiency.}  
We focus on the energy efficiency of MASA.  
MASA utilizes multiple local row buffers across
subarrays and increases the chance that an access will hit in a local
row buffer. Specifically, MASA increases the row buffer hit rate by an average
of 12.8\% across 32 benchmarks. A row buffer hit not only
has a lower access latency, but also consumes less energy, since it does not
require the power-hungry operations of activation and, to a lesser degree,
precharging. Consequently, MASA reduces the dynamic energy consumption by
18.6\% as shown in Figure~\ref{bar:row}.}

%
Our ISCA 2012 paper~\cite{salp} provides a detailed evaluation
of SALP-1, SALP-2, and MASA, including:
\begin{itemize}
\item Sensitivity studies to (1)~the number of channels (1--8), ranks (1--8),
banks (8--64), and subarrays per bank (1--128) in the memory system;
(2)~the mapping policy (row-/line-interleaved); and
(3)~an open-row or closed-policy
(Sections~9.2 and 9.3 of \cite{salp}). 

\item Multi-core results using an
application-aware memory scheduling algorithm, where we show significant performance
improvements (Section~9.3 of \cite{salp}). 

\item An analysis of the power and area overhead at
both the DRAM chip and the memory controller (Section~6 of \cite{salp}). 

\end{itemize}
%

\section{Related Work}

To our knowledge, our ISCA 2012 paper~\cite{salp} is the first to exploit the existence of subarrays
within a DRAM bank and enable their parallel operation in a cost-effective
manner. 
We propose three schemes that exploit the existence of subarrays
within DRAM banks to mitigate the negative effects of bank conflicts. Related
works propose increasing the performance and energy-efficiency of DRAM through
approaches such as DRAM module reorganization, changes to DRAM chip design, and
memory controller optimizations.  We briefly discuss these works here.

{\bf DRAM Module Reorganization.} Several prior works~\cite{ware-iccd2006,
ahn-cal2009, zheng-micro2008, ahn-taco2012} partition a DRAM rank and the DRAM
data bus into multiple rank subsets, each of which can be operated
independently. While these techniques increase parallelism, they reduce the
width of the data bus of each rank subset, leading to longer latencies to
transfer a 64~byte cache line.  Furthermore, having many rank subsets requires
a correspondingly large number of DRAM chips to compose a DRAM rank, an
assumption that does not hold in mobile DRAM systems where a rank may consist
of as few as two chips~\cite{micron-lpddr}.  Unlike these works, our mechanisms
increase memory-level parallelism~\cite{mutlu-micro2005,mutlu-hpca2003,mutlu-isca2005,mutlu-ieeemicro2003,parbs,qureshi-isca2006,cjlee-micro09} without increasing memory latency or the
number of DRAM chips.


{\bf Changes to DRAM Design.} Cached DRAM organizations, which have been widely
proposed~\cite{esdram, hart-compcon1994, hidaka-ieeemicro90, hsu-isca1993, kedem-1997, vc-sdram,
embedded-registers, dram-caching, cached-dram}, augment DRAM chips with an
additional SRAM cache that can store recently accessed data in order to
reduce memory access latency. However, these proposals increase the chip area 
and design complexity of DRAM designs. Furthermore, cached DRAM
provides parallelism only when accesses \emph{hit} in the SRAM cache, while serializing
cache misses that access the same DRAM bank. Our schemes
parallelize DRAM bank accesses while incurring significantly lower area and
logic complexity.

Fujitsu's FCRAM~\cite{sato-vlsic1998} and Micron's RLDRAM~\cite{dram-circuit-design} propose to implement shorter local bitlines 
(i.e., fewer cells per bitline) that are quickly drivable due to their 
lower capacitance in order to reduce DRAM latency. However, this significantly 
increases the DRAM die size (30-40\% for FCRAM, 40-80\% for
RLDRAM) because the large area of sense-amplifiers is
amortized over a smaller number of cells.
Hybrid memory systems can reduce the die size overhead by using
a small amount of FCRAM~\cite{sato-vlsic1998} or RLDRAM~\cite{dram-circuit-design} in conjunction
with conventional DRAM and managing which subset of the data resides in FCRAM/RLDRAM
at any given time to lower the latency of memory accesses.

A patent by Qimonda~\cite{sub-bank-patent} proposes the high-level notion of
separately addressable sub-banks, but lacks concrete mechanisms for
exploiting the independence between sub-banks. 
Yamauchi et al.\ propose the Hierarchical Multi-Bank
(HMB)~\cite{multi-bank-dram}, which parallelizes accesses to different subarrays
in a fine-grained manner. However, this scheme adds complex logic to all
subarrays. 

Udipi et al.~\cite{rethinking-dram} propose two techniques (SBA and SSA)
to lower DRAM power. In SBA, global wordlines are segmented
and controlled separately so that tiles in the horizontal direction are not
activated in lockstep, but selectively. However, this increases DRAM chip area
by 12-100\%~\cite{rethinking-dram}. SSA combines SBA with chip-granularity
rank-subsetting to achieve even higher energy savings. Both SBA and SSA
increase DRAM latency, more significantly so for SSA (due to rank-subsetting).

When transitioning from serving a write request to serving a read request, and vice
versa~\cite{staged-reads, lee2010dram, vwq-isca10}, a DRAM chip experiences bubbles in the
data bus, called the \emph{bus-turnaround penalty} ({\em tWTR} and {\em tRTW}). During
the bus turnaround penalty, Chatterjee et al.~\cite{staged-reads} propose to
internally ``prefetch'' data for subsequent read requests into extra registers
that are added to the DRAM chip.

Other works propose new DRAM designs that are capable of reducing memory latency of
conventional DRAM~\cite{chang-sigmetric16,lisa,al-dram,tl-dram,ava-dram,donghyuk-stack,donghyuk-ddma,micron-rldram3,sato-vlsic1998,hart-compcon1994,hidaka-ieeemicro90,hsu-isca1993,kedem-1997,son-isca2013,luo-dsn2014,chatterjee-micro2012,phadke-date2011,shin-hpca2014,o-isca2014,zheng-micro2008,ware-iccd2006,ahn-cal2009,ahn-taco2012}
as well as non-volatile memory~\cite{meza-weed2013,ku-ispass2013,meza-cal2012,yoon-iccd2012,qureshi-isca2009,qureshi-micro2009,lee-isca2009,lee-ieeemicro2010,lee-cacm2010}.
Previous works on bulk data transfer~\cite{gschwind-cf2006, gummaraju-pact2007,
kahle-ibmjrd2005,carter-hpca1999,zhang-ieee2001,seo-patent,intelioat,zhao-iccd2005,jiang-pact2009,seshadri2013rowclone,lu-micro2015,lisa}
and in-memory
computation~\cite{ahn-isca2015,ahn-isca2015-2,7056040,7429299,guo-wondp14,592312,
seshadri-cal2015,mai-isca2000,draper-ics2002,
seshadri-micro2015,ambit,hsieh-iccd2016,tom-isca16,
amirali-cal2016, stone-1970, fraguela-2003,375174,808425,
4115697,694774,sura-2015,zhang-2014,akin-isca2015,
babarinsa-2015,7446059,6844483,pattnaik-pact2016,vivek-chapter,pim-chapter, kim.bmc18,
boroumand.asplos18} can be used improve DRAM
bandwidth utilization and lower the number of costly data movements between CPU cores
and DRAM. \changesI{All these works can benefit from SALP as the underlying memory substrate.}




{\bf Memory Controller Optimizations.} To reduce bank conflicts and increase
row buffer locality, Zhang et al.~\cite{permutation-interleaving} propose to randomize the bank address of
memory requests by XOR hashing. Sudan et al.~\cite{micro-pages}
propose to improve row buffer locality by placing frequently-referenced data from different rows
together in the same row buffer. Both proposals can be combined with
our mechanisms to further improve parallelism and row buffer locality.

\changesI{Prior works propose memory scheduling algorithms for CPUs (e.g.,~\cite{pams, atlas, tcm, mcp, stfm-micro07, parbs, fqm,bliss,mise,ipek-isca08,morse-hpca12,pa-micro08,cjlee-micro09,vwq-isca10,lee2010dram,mutlu-podc08,ghose2013,xiong-taco16,liu-ipccc16,bliss-tpds,lavanya-asm,memattack,jishen-firm,hyoseung-rtas14,hyoseung-rts16}), GPUs (e.g.,~\cite{sms,jeong2012qos,chatterjee-sc14,medic,adwait-critical-memsched}), and other systems (e.g.,~\cite{usui-dash,usui-squash,jishen-firm})} that prioritize certain
favorable requests in the memory controller to improve system performance
and/or fairness. Subarrays expose more parallelism to the memory controller,
increasing the controller's flexibility to schedule requests. 
Our subarray-level parallelism mechanisms can be combined with many of
these schedulers to provide increased performance benefits.
\changesI{Enabling higher benefit from SALP by designing SALP-aware memory scheduling algorithms is a promising open research topic.}


\section{Significance and Long-Term Impact}

We believe SALP will have long-term impact because: {\em i)} it tackles a
critical problem, \changesI{bank conflicts and memory parallelism,} whose importance will increase in the future; and {\em ii)} \changesI{the memory substrate it provides can}
further be leveraged to enable other novel optimizations in the memory
subsystem. \changesI{In fact, as Section~\ref{sec:future-salp} shows, there has been a
significant amount of
work that built upon our ISCA 2012 paper in the past six years.}

\subsection{Trends and Opportunities in Favor of SALP}

{\bf Worsening Bank Conflicts.} Future many-core systems with large numbers of
cores and accelerators (e.g., bandwidth-hungry GPUs) will exert increasingly
\changesI{larger amount of} pressure on the memory subsystem. On the other hand, naively adding more
DRAM banks is difficult without incurring high costs, \changesI{high energy} or reduced performance.
Therefore, as more and more memory requests contend to access a limited er
of banks, bank conflicts will occur with increasing likelihood and severity.
SALP is a cost-effective mechanism to alleviate the bank conflict problem by
exploiting the existing subarrays in DRAM at low cost.

{\bf Challenges in DRAM Scaling.} DRAM process scaling is becoming more
difficult due to increased manufacturing
complexity/cost and reduced cell \changesI{reliability~\cite{kim-isca2014,mutlu-imw2013,mutlu2017rowhammer,superfri,memcon13,micron-future, itrs,kang14}}. As a result, it is critical to
examine alternative ways of improving memory performance while still maintaining
low cost. SALP is a \changesI{new cost-effective DRAM design} whose advantages
are mostly orthogonal to the advantages of DRAM process scaling. Therefore,
SALP can further improve the performance and the energy-efficiency of future
DRAM. 
In fact, a recent industry proposal to enhance the DDR standard incorporates
one of our SALP mechanisms~\cite{kang14}. \changesI{This work by Samsung and Intel 
quantitatively shows that SALP is an effective mechanism to tolerate increasing
write latencies in DRAM, corroborating the results in our ISCA 2012 paper on SALP-2.}

{\bf A Building Block for New Optimizations.} SALP enables new DRAM
optimizations that were not possible before. We discuss three potential
examples. First, exploiting subarray-level parallelism can potentially mitigate DRAM
unavailability during \changesI{refresh} by parallelizing refreshes in one subarray with
accesses to another subarray within the same bank. 
Work by Chang et al.~\cite{dsarp}, \changesI{which builds on our ISCA 2012 paper,} 
shows that such parallelization can \changesI{eliminate most} of the performance overhead of refresh.
Second, subarrays provide an
additional degree of freedom in mapping the physical address space onto
different levels of the DRAM hierarchy (channels, ranks, banks, subarrays, rows,
columns). \changesI{Thus, they enable more flexibility in performance and energy optimization 
via data mapping.} Third, DRAM can be divided among different applications (to provide
quality-of-service) at the finer-grained partitions of subarrays that are less
vulnerable to capacity and bandwidth fragmentation. \changesI{As we discuss, some research has
explored these approaches (also see Section~\ref{sec:future-salp}).} We expect \changesI{even more} 
future research will tap into these and other opportunities that can use our proposed SALP
substrate as a building block for other optimizations.

{\bf Widely Applicable Substrate.} SALP is a general-purpose substrate that is
also applicable to embedded DRAM (eDRAM)~\cite{power7edramjssc} and 3D
die-stacked DRAM (3D-DRAM), both of which consist of subarrays~\cite{loh-stack,donghyuk-stack}. For example,
eDRAM is known to be vulnerable to the write-recovery penalty~\cite{z196micro},
since it is typically used as the last-level cache and thus exposed to higher
amounts of write traffic. SALP can increase the availability of eDRAM by hiding
the write-recovery penalty. In addition, SALP may be applied to future
emerging memory technologies as long as their banks are organized
hierarchically~\cite{lee-isca2009,meza-iccd2012}, similar to how a DRAM bank consists of subarrays.

{\bf New Research Opportunities.} SALP creates new opportunities for exploiting
and enhancing the parallelism and the locality of the memory subsystem.
\begin{itemize} 
  \item {\em Enhancing Memory-Level Parallelism.} To tolerate the long latency
    of DRAM, computer architects often design mechanisms that perform multiple memory
    requests in a concurrent \changesI{manner~\cite{tomasuloOoO,mutlu-micro2005,mutlu-hpca2003,mutlu-isca2005,mutlu-ieeemicro2003,parbs,qureshi-isca2006,cjlee-micro09}}. Such efforts may become ineffective when
    requests access the same DRAM bank and, as a consequence, are not actually
    served in \changesI{parallel~\cite{parbs}}. SALP, on the other hand, parallelizes requests to
    different subarrays \emph{within the same bank}. In this regard, we believe SALP
    not only enhances previous approaches to memory-level parallelism, but also
    creates opportunities for developing new techniques that preserve
    memory-level parallelism in a subarray-aware manner.

  \item {\em Enhancing Memory Locality.} Memory access patterns that exhibit
    high locality benefit greatly from a DRAM bank's row buffer where the last
    accessed row is cached (4--8kB). While a DRAM bank has multiple row buffers
    across multiple subarrays, an existing DRAM system \changesI{exposes} only one
    row buffer at a time \changesI{in a bank} and, as a result, \changesI{is} prone to row buffer thrashing. In
    contrast, SALP allows a DRAM bank to utilize multiple row buffers
    concurrently. This enables the opportunity for new techniques that can take
    advantage of the multiple row buffers, whether they be for streaming/strided
    accesses (demand or prefetch), vector processing, or GPUs.

\end{itemize}

\subsection{Works Building on SALP}
\label{sec:future-salp}

\changesI{The introduction of the notion of subarrays and their
\sg{microarchitecture} has enabled the use of the subarrays in many works. \sg{These
include} RowClone~\cite{seshadri2013rowclone}, TL-DRAM~\cite{tl-dram},
DSARP~\cite{dsarp}, DIVA-DRAM~\cite{ava-dram}, LISA~\cite{lisa},
ChargeCache~\cite{chargecache}, Multiple Clone Row DRAM~\cite{choi-isca2015},
\sg{Ambit~\cite{ambit, seshadri-cal2015}, ERUCA~\cite{lym.hpca18},}
and other \sg{works on improving DRAM}~\cite{cream,lu-micro2015,yue-date13,son-isca2013}. Some of these
works exploit subarray level parallelism, e.g., DSARP~\cite{dsarp} reduces the overhead
of a DRAM refresh by decoupling independent subarrays from the subarray that is being refreshed.
This decoupling allows DRAM to service memory accesses while a subarray is being refreshed. 
Others make changes to subarrays to
improve an aspect, e.g., TL-DRAM~\cite{tl-dram} creates two different latency regions in a
subarray to improve DRAM latency at low cost.}

\section{Conclusion}

Our ISCA 2012 paper~\cite{salp} introduces three new mechanisms that exploit the existence of subarrays
within a DRAM bank to mitigate the performance impact of bank
conflicts. Our mechanisms are built on the key observation that
subarrays within a DRAM bank operate largely independently and have
their own row buffers. Hence, the latencies of accesses to different
subarrays within the same bank can potentially be overlapped to a
large degree. Our three mechanisms take advantage of this
fact and progressively increase the independence of operation of
subarrays by making small modifications to the DRAM chip. Our most
sophisticated scheme, MASA, enables {\em i)} multiple subarrays to be
accessed in parallel, and {\em ii)} multiple row buffers to remain
activated at the same time in different subarrays, thereby improving
both memory-level parallelism and row buffer locality. We show that
our schemes significantly improve system performance on both
single-core and multi-core systems on a variety of workloads while
incurring little ($<$0.15\%) or no area overhead in the DRAM chip. Our
techniques can also improve memory energy efficiency. 

We conclude that exploiting subarray-level parallelism in a DRAM bank can be a
promising and cost-effective method for overcoming the negative effects of DRAM
bank conflicts, without paying the large cost of increasing the number of banks
in the DRAM system. \changesI{Significant recent work has built upon our ISCA
2012 paper, and we expect \sg{many other new works} can exploit the new substrate we have
enabled to achieve even bigger goals and higher benefits.}


\section*{Acknowledgments}

\changesI{We thank Rachata Ausavarungnirun and Saugata Ghose for their dedicated effort in the preparation of this article.}
Many thanks to Uksong Kang, Hak-soo Yu, Churoo Park,
Jung-Bae Lee, and Joo Sun Choi from Samsung for their helpful
comments. We thank the anonymous reviewers for their
feedback. We gratefully acknowledge members of the SAFARI
group for feedback and for the stimulating intellectual environment
they provide. We acknowledge the generous support of
AMD, Intel, Oracle, and Samsung. This research was also partially
supported by grants from NSF (CAREER Award CCF-
0953246), GSRC, and Intel ARO Memory Hierarchy Program.

{
\bibliographystyle{IEEEtranS}
\bibliography{paper,tldram}
}

\end{document}